
\magnification 1200
\hsize 16. true cm             
\baselineskip = 6truemm        

\def\parn{\par\noindent}

\def\PR{{\it Phys.Rev. }}

\def\ref#1{[#1]}
\def\oo{\infty}

\def\Li{{\rm Li}}
\def\qk{{\hat q \cdot \hat k}}
\def\QK#1{{\hat q \cdot \hat k_#1}}
\def\KK{{\hat k_1 \cdot \hat k_2}}
\def\dk{{(k_1-k_2)^2}}
\def\rabc{{R(k_1^2,k_2^2,\dk)}}
\def\rqkm#1{{R(q^2,k_#1^2,-m_#1^2)}}
\def\rqam#1{{R(q^2,k_#1^2,-m_#1^2)}}
\def\rqbm#1{{R(q^2,k_#1^2,-m_#1^2)}}
\def\RD{\sqrt{\Delta}}

\font\fontA=cmbx10 scaled\magstep1
\def\d${$ \displaystyle }

\rightline{\bf DFUB 94-01}
\rightline{24 February 1994}
\vskip 20truemm
\centerline{\bf Hyperspherical integration and the triple-cross vertex graphs}
\par
\vskip 20truemm
\centerline{S.Laporta
\footnote {$^1$}{ {E-mail: {\tt laporta@bo.infn.it} }}  }
\vskip 20truemm
\centerline{\it Dipartimento di Fisica, Universit\`a di Bologna,}
\centerline{\it and INFN, Sezione di Bologna,}
\centerline{\it Via Irnerio 46, I-40126 Bologna, Italy}
\vskip 1 truecm
\parn
\vskip 20truemm
\centerline{\bf Abstract. }  \par
We have extended the hyperspherical variables method
to the analytical calculation of the angular integral of the box graph.
We discuss the applications of our results to the
analytical calculation of the QED contribution to the electron $g$-2
of the set of three-loop triple-cross vertex graphs.
\vskip 2truemm
\centerline {
----------------------------------------------------------------------------}
\parn
PACS 12.20.Ds - Specific calculations and limits
                of quantum electrodynamics.
\parn
PACS 12.90\phantom{.Ds} - Miscellaneous theoretical ideas and models.
\vfill\eject
%
\parn {\bf 1.}\hskip 1 truemm
New planned high precision
experimental determinations of the electron $g$-2
demand correspondent higher precision calculations of
QED theoretical contributions.
One-loop, two-loop  and almost all the three-loop QED contributions
to the electron $g$-2 are known in analytical form.
The analytical calculations of the known three-loop contributions
relied on two approaches: the hyperspherical variables method [1][2]
and the dispersion relations method [2][3],
the choice being dependent on the topology of the graphs.
A hybrid dispersive and hyperspherical approach
turned out to be more convenient for some troublesome sets of graphs
[4][5].

There are two sets of three-loop graphs whose contributions
are still not known in analytical form, but only in numerical form.
Analytical calculation of one of them,
the set of graph called ``corner-ladder'', is in progress [6].
On the contrary, none of the above mentioned analytical approaches
seems to be suitable for the analytical calculation of the last
remaining
set of (not planar) graphs  called ``triple-cross'' (see Fig.1).
The application of the dispersion relations method to the whole graph
seems difficult because of the high number
of cut graphs and the complexity of the calculations involved.
On the other hand,
the hyperspherical variables method (in the formulation of ref.[1])
cannot be applied to box graphs or not planar vertex graphs at all.
No useful way of combining both methods has been found so far.

In this paper we present an extension of the hyperspherical variables
method to the box graph and we discuss the application of our results
to the calculation of a contribution of a typical not planar
``triple-cross'' graph.
\parn {\bf 2.}\hskip 1 truemm
%
Let us summarize the main steps of the analytical calculation
of a Feynman integral by means of the hyperspherical variables approach [1]:
\par
1) The Feynman integral is Wick-rotated to an Euclidean integral;
   so doing, the external momenta become all spacelike, and the result
   of the integration must be eventually continued to the timelike region.
  \par
2) Four-dimensional hyperspherical polar coordinates are introduced
   for each loop momentum.
  \par
3) The angular integrations are analytically performed.
  \par
4) The remaining radial integrations (one for each loop) are performed
   analytically.
   \par
Obviously the points 1) and 2) do not depend on the particular kind of graph
considered.
The point 3) is more critical;  an elegant and fast method for performing
the angular integrations is presented in ref.[1]:
the denominators are expanded in Gegenbauer polynomials
(related to four-dimensional hyperspherical harmonics),
the angular integrals are performed using orthogonality relations and
the infinite series so obtained is summed up analytically.
The feasibility of the subsequent radial integration is related to
the analytical form of the results of the angular integrals.
As an example, let us consider  the simple 4-dimensional angular integral
of a scalar bubble-loop graph
$$ \int {d\Omega_q \over {(q-k)^2+m^2}} \ , \eqno (1) $$
where $q$ and $k$ are vectors in a 4-dimensional Euclidean space.
The denominator of eq.(1) is expanded in Gegenbauer polynomials $C^1_n(\qk)$:
$$  {1\over {(q-k)^2+m^2}}
  =  {1\over |q||k|} {1\over {1-2Z \qk +Z^2}}
  =  {1\over |q||k|} \sum_{n=0}^\oo Z^{n+1} C_n^1\left(\qk\right) \ ,
\eqno(2)$$
where
$$ Z={{q^2+k^2+m^2-R(q^2,k^2,-m^2)}\over{2|q||k|}} \ ,$$
$$ R(x,y,z)=\sqrt{x^2+y^2+z^2-2xy-2xz-2yz} \  ;$$
the angular integrations are performed by using the relation
$$ \int d\Omega_q  \; C^1_n(\qk) =  {2\pi^2} \delta_{0n} \ ,$$
so that the summation of the infinite series (2) becomes trivial and
$$ \int {d\Omega_q \over {(q-k)^2+m^2}} = {2\pi^2}  {Z\over |q||k|} \ .
\eqno (3) $$

Successful application of the above described method to
more complex angular integrals depends strongly on the momentum routing
and of topology of the correspondent graphs;
in fact, these conditions must be satisfied [1]:
\par
i)  The momentum of each line of the graph must be a linear combination
    of only two distinct momenta, loop or external;
    for every planar vertex graph one can find
    a momentum routing which satisfies
    such a condition, while
    in the case of the not planar ``triple-cross'' vertex graphs
    this is not possible.
\par
ii) At least one vector must not appear in more than two Gegenbauer
    polynomials,
    as tables of integrals of product of three Gegenbauer polynomials
    containing
    the same vector are not known;
    the box graph and all graphs made up only of box subgraphs
    do not satisfy this condition.

Let us now consider the angular integral containing two denominators
$$ \int {d\Omega_q \over {[(q-k_1)^2+m_1^2] [(q-k_2)^2+m_2^2] } }
                                  \eqno (4)  $$
which is the hyperspherical angular integral of a triangle loop;
this integral can be worked out in the same way of previous integral (1);
in fact, applying the expansion (2) to both denominators, one works out
the double series
$$ \int {d\Omega_q \over {[(q-k_1)^2+m_1^2] [(q-k_2)^2+m_2^2] } }
  =  {1\over q^2|k_1||k_2|}
\times $$
$$ \qquad \qquad
   \times \sum_{m=0}^\oo \sum_{n=0}^\oo
 Z_1^{m+1} Z_2^{n+1} C_m^1\left(\QK1\right) C_n^1\left(\QK2\right) \ ,
 \eqno (5) $$
$$ Z_i={{q^2+k_i^2+m_i^2-R(q^2,k_i^2,-m_i^2)}\over{2|q||k_i|}} \ ,$$
which, by using the orthogonality properties of the hyperspherical harmonics
$$ \int C^1_m(\QK1) \; C^1_n(\QK2) \; d\Omega_q = {2\pi^2}\
                      {\delta_{mn}\over {n+1}} C^1_n(\KK) \;, \eqno(6) $$
is reduced to one single series whose sum contains two logarithms of complex
arguments containing $Z_1$ and $Z_2$;  we do not show this result here.

On the contrary, we underline a simple but important fact:
the angular integrals (1) and (4) can
be calculated directly integrating over the three hyperspherical
 angles,
avoiding the use of the expansion (2), and writing
$$\int {d\Omega_q}=\int_0^\pi   \sin^2\theta_1 d\theta_1
                   \int_0^\pi   \sin  \theta_2 d\theta_2
                   \int_0^{2\pi}               d\theta_3 \ ;\eqno(7) $$
the calculations become rather cumbersome, but, as it will be shown
later, they are still feasible when the expansion (2) cannot be applied.

Let us now calculate the integral (4)
in analytical form using the decomposition (7).
We combine the denominators by using one Feynman parameter,
the simple angular integration is done by using
the 1-denominator result (3),
and the subsequent integration over the Feynman parameter is
straightforward.

The analytical result contains logarithmic functions of
rather complicated arguments that we schematize as
$$\log { {\alpha +\beta R_1 +\gamma R_2 +\delta R_1 R_2}\over
         {\alpha +\beta R_1 -\gamma R_2 -\delta R_1 R_2} }=
         \log {{N(R_1,R_2)}\over {N(R_1,-R_2)}} \ , \eqno(8) $$
where $\alpha$, $\beta$, $\gamma$, $\delta$ are polynomials
and $R_1$ and $R_2$ are square roots of polynomials of the radial
 variables;
we note that the logarithms appearing in the result obtained using
the expansion in Gegenbauer polynomials
have a similar complicated form.
In both cases, for continuing the analytical integration
 over the radial variable $|q|$
it is essential to decompose these logarithmic functions [5],
by separating the various combinations of square roots $R_i$
so that only functions of the form
         $$ \log {{a+b R_1}     \over {a-b R_1}}\ ,
            \log {{c+d R_2}     \over {c-d R_2}}\ ,
            \log {{e+f R_1 R_2} \over {e-f R_1 R_2}}$$
      appear.
In fact, by observing that the product
$$ N(R_1,R_2) N(-R_1,R_2) N(R_1,-R_2) N(-R_1,-R_2,)$$
is a polynomial squared and that
$$ N(R_1,R_2) N(-R_1, R_2) = P_1 (c+d R_2)^2 \ ,$$
$$ N(R_1,R_2) N( R_1,-R_2) = P_2 (a+b R_1)^2 \ ,$$
$$ N(R_1,R_2) N(-R_1,-R_2) = P_3 (e+f R_1 R_2)^2 \ ,$$
($P_i$ are polynomials), we can rewrite eq.(8) as
$$           \log {{N(R_1,R_2)}\over {N(R_1,-R_2)}}=
  {1\over 2} \log
               {{N(R_1,R_2) N(-R_1, R_2)} \over {N(R_1,-R_2) N(-R_1,-R_2)}}
               {{N(R_1,R_2) N(-R_1,-R_2)} \over {N(R_1,-R_2) N(-R_1, R_2)}}
    \ , $$
so that
$$           \log {{N(R_1,R_2)}\over {N(R_1,-R_2)}}=
   \log {{c+d R_2}\over {c-d R_2}}
  +\log {{e+f R_1 R_2}\over {e-f R_1 R_2}}  \ . \eqno (9) $$

Applying the decomposition here exemplified
to all the logarithmic functions,
the analytical expression of the integral (4) takes the simpler form:
$$ \int {d\Omega_q \over {[(q-k_1)^2+m_1^2] [(q-k_2)^2+m_2^2] }}  =
{\pi^2\over {q^2 \rabc}} \times$$
$$\eqalign {
\times\Biggl\{&\log {{C_1+\rabc\rqam1}\over {C_1-\rabc\rqam1}} \cr
            +&\log {{C_2+\rabc\rqbm2}\over {C_2-\rabc\rqbm2}}  \cr
            +&\log {{C_3+\rabc}      \over {C_3-\rabc}}
                 \Biggr\}  \cr
} \eqno(10) $$
where

$$
\eqalign{
C_1 =&(k_2^2-k_1^2-\dk)q^2 + k_1^4 +m_1^2(k_1^2+k_2^2-\dk) \cr
                                  & - k_1^2(k_2^2+\dk+2m_2^2) \ ,\cr
C_2 =&(k_1^2-k_2^2-\dk)q^2 + k_2^4 +m_2^2(k_1^2+k_2^2-\dk) \cr
                                  & - k_2^2(k_1^2+\dk+2m_1^2) \ ,\cr
C_3 =&\dk-k_1^2-k_2^2 \ . \cr
 } \eqno(11)$$

The symmetry of eq.(10)
under the exchange $(k_1,m_1)\leftrightarrow (k_2,m_2)$ is evident;
we underline that eq.(10) has a form simpler than the result obtained
using the expansion (2), which contains more complicated logarithms of
the kind (8).

Let us continue our analysis,
considering this time the angular integral containing three denominators
$$ \int {d\Omega_q \over {[(q-k_1)^2+m_1^2]
                          [(q-k_2)^2+m_2^2]
                          [(q-k_3)^2+m_3^2] } } \eqno (12) $$
which is the hyperspherical angular part of the amplitude of the
box graph shown in
Fig.2.
Using again the expansion (2), we work out a triple series
containing the angular integrals
$$ \int           C^1_m(\QK1) \;
                  C^1_n(\QK2) \;
                  C^1_p(\QK3) \; d\Omega_q \ \;. \eqno(13) $$
The analytical expressions of above integrals are not known;
the known summation formulae for
hyperspherical harmonics,
which allows to work out eq.(6),
are not useful in this case;
the reason of this behaviour will be more clear in the following.
Therefore the integral (12) cannot be calculated in analytical form
using the method shown in ref.[1].
Thus, following the above described two-denominators example,
we combine the three
denominators of eq.(12) introducing two  Feynman parameters,
we perform the angular integration using the formula
$$ \int {d\Omega_q \over {\left[(q-k)^2+m^2\right]^3}} =
{{2\pi^2}\over {R^3(q^2,k^2,-m^2)} } \eqno(14) $$
and the subsequent integrations over the Feynman parameters which are
straightforward but rather cumbersome;
the result is a sum of four logarithmic functions
of asymmetrical huge arguments which are decomposed with the same mechanism
of eq.(8-9) and recombined into three logarithms of simpler arguments.
We expect the result to have a structure similar to that of eq.(10).
In fact one obtains
$$ \int {d\Omega_q \over {[(q-k_1)^2+m_1^2]
                          [(q-k_2)^2+m_2^2]
                          [(q-k_3)^2+m_3^2] }}=
{{\pi^2}\over{q^2\RD}} \times $$
$$ \times
\Biggl\{ \log {{C_1+\rqkm1\RD}\over {C_1-\rqkm1\RD}}
        +\log {{C_2+\rqkm2\RD}\over {C_2-\rqkm2\RD}} $$
$$+\log {{C_3+\rqkm3\RD}\over {C_3-\rqkm3\RD}} \Biggr\} \eqno(15) $$
where  $\Delta$ is the polynomial of degree 2 in $q^2$
$$ \Delta=\alpha_2 (q^2)^2 +\alpha_1 q^2 +\alpha_0 $$
with the coefficients
$$ \eqalign {
  \alpha_2=&R^2(a_1,a_2,a_3) \cr
  \alpha_1=&
   2 u_1  (2 a_1 b_1 - a_1 b_2 - a_1 b_3 + a_2 b_3
                      + a_3 b_2 - a_2 b_2 - a_3 b_3 )
\cr
&
  +  2 (u_1-b_1)a_1( a_1 - a_2 - a_3)
        - {4\over3} a_1 a_2 a_3
        \quad \left( +  {\rm 2\  cyclic\  permutations}\right) \cr
   \alpha_0=&
     u_1^2   (a_1^2 - 2 a_1 b_2 - 2 a_1 b_3 - 2 b_2 b_3 + b_2^2 + b_3^2 ) \cr
  & + 2 u_1 u_2   (  - a_1 a_2 + a_1 b_1 + a_1 b_3 + a_2 b_2 + a_2 b_3
   - 2 a_3 b_3 - b_1 b_2 + b_1 b_3 + b_2 b_3 - b_3^2 ) \cr
&  + 2 u_1   ( a_1^2 b_1 - a_1 a_2 b_2 - a_1 a_3 b_3 - a_1 b_1 b_2 - a_1 b_1
b_3
       + 2 a_1 b_2 b_3
        + a_2 b_2^2 - a_2 b_2 b_3
    \cr
&  - a_3 b_2 b_3 + a_3 b_3^2 )
        + a_1^2 b_1^2 - 2 a_1 a_2 b_1 b_2
        \quad \left( +  {\rm 2\  cyclic\  permutations}\right) \cr
}$$
where
$a_1=(k_2-k_3)^2$, $a_2=(k_1-k_3)^2$, $a_3=(k_1-k_2)^2$,
$ b_i=k_i^2$ and $u_i=m_i^2$
(we show a partial list of the expressions of
 $\alpha_0$ and $\alpha_1$ with the indication that the complete
 expressions
 must be obtained summing up also the two cyclic permutations
 of the indices 1,2 and 3 of the shown portions).

The arguments of the logarithms contain
$$ \eqalign {
    C_1=& \beta_2 q_2^2 +\beta_1 q_2 +\beta_0 \cr
    \beta_2 =&  a_1 - a_2 - a_3 \cr
    \beta_1 =&  u_1   ( 2 a_1 - a_2 - a_3 + 2 b_1 - b_2 - b_3 )
       + u_2   (  - a_2 - b_1 + b_3 )   + u_3   (  - a_3 - b_1 + b_2 )
\cr
&
    - 2 a_1 b_1 - 2 a_2 a_3 + a_2 b_1 + a_2 b_2 + a_3 b_1 + a_3 b_3 \cr
    \beta_0 =&
         u_1^2      ( a_1 - b_2 - b_3 )
       + u_1 u_2   (  - a_2 + b_1 + b_3 )   + u_1 u_3 (  - a_3 + b_1 + b_2 )
\cr
&
       + u_2 b_1   (  - a_2 + b_1 - b_3 )
       + u_3 b_1 (  - a_3 + b_1 - b_2 )
       + u_2 u_3   (  - 2 b_1 )   \cr
&      + u_1   ( 2 a_1 b_1 - a_2 b_2 -   a_3 b_3
                 - b_1 b_2 - b_1 b_3 + 2 b_2 b_3 )
       + a_1 b_1^2 - a_2 b_1 b_2 - a_3 b_1 b_3
      \cr
} $$
and
$$
\eqalign{
 C_2=& C_1 ({\rm with\  the\ cyclic\ permutation\ } 1\rightarrow2,
                                                  \;2\rightarrow3,
                                                  \;3\rightarrow1 )\cr
 C_3=& C_1 ({\rm with\ the\ cyclic\ permutation\ } 1\rightarrow3,
                                                 \;2\rightarrow1,
                                                 \;3\rightarrow2 )
                                                           \ .\cr
} $$

We note that:
\par
1) Due to the effect of decomposition (8-9),
   eq.(15) has a form such that the symmetry under the permutations
  of the indices 1,2 and 3 is evident at sight.
\par
2) Comparing eq.(15) and eq.(10) we observe that
   the square root $\RD$ plays a role analogous to that of
   $\rabc$ in eq.(10), but it has a more complex structure,
   not reducible
   to  one single square root of the kind $R(x,y,z)$.
This fact explains why the expansion in hyperspherical harmonics
   fails to obtain our result eq.(15):
   if the integrals of
   three Gegenbauer polynomials (13) were known in analytical form,
   we would be able to perform
   the angular integration of the product of the
   expansions of the three denominators of eq.(12);
   but the result
   would be the expansion of our result (15)
   in Gegenbauer polynomials, surely very complicated,
   whose summation in analytical form to (15)
   would be almost hopeless.
\par
3) The angular integral of the box graph (15)
   can be used as a ``building block'' in order to
   work out more complex angular integrals.
\vskip 6 truemm \parn {\fontA \bf 3.}
In this section we apply the result of the previous section
to the analytical calculation of the contribution to electron $g$-2
of a typical ``triple-cross'' graph.

We consider the scalar amplitude $I$ of
the not planar self-energy graph obtained from
the vertex graph of Fig.1, disregarding the external photon interaction
and the numerators of the Feynman integral.

Following the section (2), we decompose the integral
in radial and angular parts:
$$I=\int{{dk_1^2}\over 2}
    \int{{k_2^2\;dk_2^2}\over 2}
    \int{{q^2\;dq^2}\over {2(q^2+1)}} J(k_1^2,k_2^2,q^2,p^2) $$
where $ J(k_1^2,k_2^2,q^2,p^2) $ is the angular integral
$$ J(k_1^2,k_2^2,q^2,p^2)=
    \int{{d\Omega_{k_2}}\over {(p-k_2)^2}}
    \int{{d\Omega_{q}}\over {(q-k_2)^2 \left[(p+q-k_2)^2+1\right]}}
\times $$
$$ \times
    \int{{d\Omega_{k_1}}\over {(k_1-p)^2  \;
                              \left[(k_1-k_2)^2+1\right] \;
                              \left[(k_1-q)^2+1\right] }     } \
                                        \eqno (16) $$
and the electron mass has been set to 1.
We underline that this graph is not planar and,
from a topological point of view,
it is made up of three box loops with one or two common sides;
the momentum routing has been chosen in such a way that only one
denominator be a linear combination of three momenta
(this fact is unavoidable because of the not planarity of the graph).
As we have shown in the previous section,
the expansion of the denominators is not useful for evaluating this angular
integral in analytical form,
therefore we apply eq.(15) to the angular integration over $k_1$,
so we find:
$$ J(k_1^2,k_2^2,q^2)=
    {\pi^2\over q^2}\int{{d\Omega_{k_2}}\over {(p-k_2)^2}}
    \int{{d\Omega_{q}}\over {(q-k_2)^2 \left[(p+q-k_2)^2+1\right]}}
    \sum_{i=1}^3 {\log F_i\over{\RD} } \ ;$$
the arguments of the logarithms contain the square roots
$R(k_1^2,p^2,-1)$,
$R(k_1^2,k_2^2,-1)$,
$R(k_1,q^2,-1)$,
$\RD$,
whose only $\RD$ depends on the angles among $p$, $q$ and $k_2$.
We introduce hyperspherical angles for the remaining angular
integrations

$$  \int{d\Omega_{k_2}} \int{d\Omega_{q}}=
  \int_0^\pi  4\pi \sin^2\theta_1 d\theta_1 \;
  \int_0^\pi  2\pi \sin^2\phi_1 d\phi_1 \;
  \int_0^{2\pi}      \sin\phi_2 d\phi_2 \  $$
where $\theta_1$ and $\phi_1$ are respectively
the angle between $p$ and $k_2$
and between $q$ and $k_2$, and $\phi_2$ is the azimuthal
angle between $p$ and $q$;
the result of the integration over $\phi_2$ is schematized as
$$ J(k_1^2,k_2^2,q^2)= \pi^4
 \int_0^\pi {\sin^2\theta_1 \over {p^2+k_2^2-2|p||k_2| \cos\theta_1}} d\theta_1
 \int_0^\pi {\sin^2\phi_1   \over {q^2+k_2^2-2|q||k_2| \cos\phi_1}} d\phi_1
\times $$
$$ \times
    {1\over \sqrt{\Delta_0}} \sum_{i=1}^{30} \Li_2(G_i) \eqno(17) $$
where $\Delta_0$ is a polynomial of degree 2 in $\cos \theta_1$ and
$\cos\phi_1$.
A careful analysis of the structure of the square root
$\sqrt {\Delta_0}$ and of the zeroes of the two denominators of eq.(17)
containing $\cos \theta_1$ and $\cos \phi_1$
shows that the angular integration over $\theta_1$ is of elliptic kind,
therefore the angular integral (16) cannot be reduced in closed analytical
form. On the contrary we have worked out a integral representation.
This not expected behaviour persists even using different momentum routings;
it is probably due to the not planarity of considered graph.

However, this fact does not stop our analytical calculation.
We found that the total angular integral of the
two-loop self-energy graph with two photon crossed
is expressed
by a elliptic integral involving logarithms obtained from eq.(8)
if one uses a particular momentum routing;
in this case an exchange between the radial integrations
and the remaining angular integral avoids the ellipticity and
allows one to complete the analytical calculation.

Considering our three-loop result (17), one finds that
$\Delta_0$ is a polynomial of degree two in $q^2$,
so that an exchange between the integrations over $\theta_1$ and $q^2$
allows one to continue the calculation.
Work is in progress to extend this mechanism to the remaining
radial integrations.

\vskip 2 truemm

All the algebraic manipulations were carried out through the symbolic
manipulation program ASHMEDAI [7].

\par
\par
\vfill\eject
\vskip 20 truemm
{\bf References}
\vskip 10 truemm

\item{[1]}  M.J. Levine and R.Roskies, \PR D {\bf 9}, 421 (1974).
\item{[2]}  R.Z.Roskies, E.Remiddi and M.Levine in
            {\it Quantum Electrodynamics},
            editor \par T.Kinoshita,
            (World Scientific, Singapore, 1990), p.162.
\item{[3]}  R.Barbieri, J.A.Mignaco and E.Remiddi,
              {\it Nuovo Cimento A}, {\bf 11}, 824-865 (1972).
\item{[4]}  M.J. Levine, E. Remiddi and R. Roskies, \PR D {\bf 20},
            2068 (1979); \par
            S. Laporta and E.Remiddi, {\it Phys.Lett.} {\bf B}265, 182 (1991).
\item{[5]}  S. Laporta and E.Remiddi, {\it Phys.Lett.} {\bf B}301, 440 (1993).
\par
\item{[6]} M. Caffo, E. Remiddi and S. Turrini,
              {\it Nuovo Cimento A}, {\bf 2}, 220 (1984).
\item{[7]} M. J. Levine, U.S. AEC Report No. CAR-882-25 (1971)
                                                    (unpublished).
\parn
\vfill\eject
{\bf Figure captions}
\vskip 12 truemm
\parn
Fig.1: An example of triple-cross vertex graph. The cross indicates
       the external photon interaction.
\parn
Fig.2: The box graph.
\parn
\vskip 12 truemm

\bye